\documentclass[intlimits,twoside,a4paper]{article}

\usepackage{amsmath,amssymb}
\usepackage{graphicx}
\usepackage{color}
\usepackage{wrapfig}

\usepackage[T2A]{fontenc}
\usepackage[cp1251]{inputenc}
\usepackage[eqsecnum]{cmpj2}

\issue{2014}{17}{1}{13502}
\doinumber{10.5488/CMP.17.13502}

\title[Long-range interaction between dust grains in plasma]{Long-range interaction between dust grains in plasma}

\author[D.Yu. Mishagli]{D.Yu. Mishagli\thanks{E-mail: mishagli@onu.edu.ua}}
\addresses{\addr{ad1}Department of Theoretical Physics, Faculty of Physics, Mechnicov National University,\\
2 Dvoryanska St., 65026 Odessa, Ukraine
\addr{ad2} Institut f\"ur Theoretische Physik, Universit\"at Leipzig,
10/11 Augustusplatz, 04109 Leipzig, Germany\thanks{Present address}}

\authorcopyright{D.Yu. Mishagli, 2014}

\date{Received March 21, 2013, in final form November 7, 2013}
\begin{document}

\maketitle

\begin{abstract}
The nature of long-range interactions between dust grains in plasma is discussed. The dust grain interaction potential within a cell model of dusty plasma is introduced. The attractive part of intergrain potential is described by multipole interaction between two electro-neutral cells. This allowed us to draw an analogy with molecular liquids where the attraction between molecules is determined by dispersion forces. The main ideas of the fluctuation theory for electrostatic field in a cell model are formulated, and the dominating contribution to the attractive part of intergrain potential is obtained.
\pacs 52.25.Gj, 52.27.Lw
\end{abstract}

\section{Introduction}
Dusty (complex) plasma consists of a weakly ionized gas (plasma) and charged submicron- and micron-sized particles (grains). It represents a new type of soft matter. It is an interdisciplinary field of research: geophysics, geology, meteorology, ecology, planetary science, different applications in technology~--- dusty plasmas can bring new results to all these directions. Progress in the research of dusty plasma properties is documented in resent monographs~\cite{bouchoule_book,shukla_book,vladimirov_book,liberman_book,fortov_book,betz_book,tsytovich_book,forfill_book,bonitz_book} and review articles~\cite{tsyt1_ufn,smirnov_ufn,shukla_plas,fortov_ufn,shukla_rev,morfill_rev,chaud}. However, there are still problems to be solved.

Dust grains being highly charged (up to $10^6$ elementary charges per grain) substantially affect the properties of the whole system. Dust grains can be either positively or negatively charged. The processes of charging are very important in the theory of dusty plasma. However, we will not consider them in the present paper (for this purpose see the above mentioned references and the recent article by the author~\cite{mish}). Under certain conditions, dust grains can form various condensed (``plasma crystal'' and ``plasma liquid'') or gaseous phases depending on the relative strengths of the intergrain interaction. In a theoretical research, the problem of describing an intergrain interaction and phase transition in a subsystem of dust grains holds an important position.

The forces that govern the dust grains do not correspond to a direct Coulomb interaction and are long-range  ones (see e.g., work~\cite{tsyt2_ufn}). Thus, in the present paper the model intergrain potential is proposed that predicts both the repulsion (at small distances) and the attraction (at large distances) between the dust grains. The cell approach to dusty plasma discussed in work~\cite{mish} plays an essential role in our construction. It allowed us: (i) to create a simple theory of charge fluctuations and (ii) to study a long-range interaction between the dust grains. The approach presented can be used to describe the interaction effects in similar systems such as the mixture of ionic and nonionic liquids \cite{aerov}, where spherical clusters occur.

The article is built as follows. In section~\ref{sec:2}, basic statements of the cell model and some results obtained in~\cite{mish} are reviewed. In section~\ref{sec:3}, we present a model intergrain interaction potential. In order to establish the main contribution to the attractive part of intergrain potential, we consider an electro-neutral cell in the external electric field in section~\ref{sec:4}, where the {dipole} polarizability of a cell is obtained. Then, in section~\ref{sec:5} we present the principles of construction of the fluctuation electrostatic field inside and outside a cell and obtain fluctuation multipole moments of a cell. Moreover, the fluctuation dipole moment is obtained and the main contribution to the attractive part of the intergrain potential is presented in the explicit form. A brief discussion of the obtained results is presented in the concluding section~\ref{sec:6}. In appendix~\ref{sec:appendix_a}, we get an expression for the potential of the fluctuation electrostatic field inside a cell. In appendix~\ref{sec:appendix_b}, the energy of the fluctuation electrostatic field is obtained.

\section{The cell model of complex plasma}\label{sec:2}

Let us consider a system of charged dust grains of the same radii $r_{\text p}$ and the emitted electrons. The dust grains have an average charge $Ze$ ($e$ is the electron charge). The system is in a thermal equilibrium (thermal plasma). We assume that the complex plasma can be represented as the collection of electro-neutral cells due to its electro-neutrality. Each cell contains only one dust grain. In the mean field approximation, the cells should have a spherical form of the radius
\begin{equation}\label{eq:cell}
r_{\text c} = \frac12 \left( \frac{3}{4 \pi n_{\text p}} \right)^{1/3},
\end{equation}
where $n_{\text p}$ is an average dust grain density. Note, that such an approach is applicable only for describing the equilibrium thermodynamic properties of a dusty plasma.

The electro-neutrality condition for a cell is as follows:
\begin{equation}\label{eq:neutrality_condition}
Z e + \int \rho (\mathbf r) \rd\mathbf r = 0,
\end{equation}
where $\rho (\mathbf r)$ is the volume-charge density, the distance $r$ is reckoned from the center of a grain. The integration occurs over the area occupied by electrons inside a cell. Note, that in the absence of an external electric field, the distributions of an electric charge $\rho (\mathbf r)$ and electric potential $\phi(\mathbf r)$ have spherical symmetry: $\rho (\mathbf r) \Rightarrow \rho(r)$, $\phi (\mathbf r) \Rightarrow \phi_0(r)$. The electrostatic field distribution is described in the self-consistent field approximation: the potential $\phi(\mathbf r)$ satisfies the Poisson equation, in which the charge density $\rho(\mathbf r)$ is determined by the Boltzmann distribution.

Its use is justified by the inequality
\begin{equation}\label{eq:inequality}
\tau_{\text c} \ll \tau_{\ast}\,,
\end{equation}
where $\tau_{\text c}$ is the time required for the formation of an electro-neutral cell around the grain, $\tau_{\ast}$ is the characteristic macroscopic relaxation time for a system. This inequality expresses the fact that the electron mobility substantially exceeds the mobility of other plasma components.

Reference \cite{mish} discusses the system of identical dust grains of the radius $r_{\text p}$ with the mean charge $Ze$, which are in equilibrium with the emitted electrons, without an external electric field. The problem of proper boundary conditions for such a model is also considered therein. It is shown that (i) setting the electrostatic potential $\phi$ equal to zero on the surface of a cell and (ii) connecting the electrostatic field strength on the surface of a grain with its average charge are sufficient conditions for a full description of a system:
\begin{equation}\label{eq:boundary_cond_1}
\left\{
\begin{aligned}
&     \phi_0 (r_{\text c}) = 0,
\\
  &  \left. \frac{\partial \phi_0 (r)}{\partial r} \right|_{r = r_{\text p}} = - 4\pi\sigma,
\end{aligned}
\right.
\end{equation}
where $\sigma$ is the average surface charge density on a dust grain. Note, that there is no electric field outside a cell due to the Gauss law.

For a further analysis in~\cite{mish}, the dimensionless variables
\begin{equation}\label{eq:dimless_varz}
\tilde r = \frac{r}{r_{\text p}}, \qquad \varsigma = \frac{r_{\text c}}{r_{\text p}}, \qquad \psi_0(\tilde r) = 1 + \frac{e \phi_0(\tilde r)}{kT}
\end{equation}
were used. Here, $\varsigma$ and $\psi_0(\tilde r)$ are, respectively, the dimensionless radius of a cell and the dimensionless potential of the electrostatic field inside a cell, and $k$ is the Boltzmann constant. Note, that $\tilde r \in [1,\varsigma]$. The solution of the linearized Poisson equation ($e\phi_0(r) \ll kT$) satisfying the boundary conditions $\eqref{eq:boundary_cond_1}$ is as follows:
\begin{equation}\label{eq:potential_psi_0}
    \psi_0(\tilde r) = \frac{1}{\tilde r}
    \frac{(Z/Z_0) \lambda \sinh \frac{\varsigma - \tilde r}{\lambda} + \varsigma \left( \lambda \sinh \frac{\tilde r - 1}{\lambda} + \cosh
    \frac{\tilde r - 1}{\lambda} \right)}
         {\lambda \sinh \frac{\varsigma - 1}{\lambda} + \cosh \frac{\varsigma - 1}{\lambda}}\,.
\end{equation}
Here, $Z_0 = kTr_{\text p}/e^2$ (typical values of $Z/Z_0 \sim 2-4$~\cite{morfill_rev}) and $\lambda$ is the dimensionless Debye radius defined~as
\begin{equation}\label{eq:deb_len}
\lambda = \frac{r_{\text D}}{r_{\text p}}\,, \qquad r_{\text D} = \sqrt{\frac{kT}{4 \pi e^2 n_{\text e0}}}\,,
\end{equation}
where $n_{\text e0}$ is the mean density of the emitted electrons for $\phi_0(r) = 0$, i.e., on the boundary of a cell (for $r=r_{\text c}$ or, in dimensionless form, $\tilde r = \varsigma$). The buffer gas is not taken into account here. Note, that in the  presented cell model dust grains do not take part in the screening.

The equation \eqref{eq:neutrality_condition} allows one to establish the dependence of Debye radius $\lambda$ on the average dust grain charge $Z$, as it is also shown in~\cite{mish}. The dependence $\lambda=\lambda(Z)$ is set as a root of the equation
\begin{equation}\label{eq:neutral_eq_3}
\frac{Z}{Z_0} = \frac{1}{\lambda} \left[ (\varsigma - \lambda^2) \sinh \textstyle \frac{\varsigma - 1}{\lambda} + \lambda (\varsigma - 1) \cosh \frac{\varsigma - 1}{\lambda} \right].
\end{equation}

In the mean-field approximation described above, there is no interaction between the cells.  However, as it is shown below, the charge fluctuations (beyond the mean-field cell approach) can lead to the interaction between two electro-neutral cells.

\section{Averaged potential of multipole interaction between dust grains}\label{sec:3}

The equilibrium value of the dust grain charge is mainly violated by the thermal electron motion. Therefore, the fluctuation electric multipole moments of cells occur and generate long-range electric fields.

The electric field of one cell acts on its neighbours and generates interaction effects, which are similar to a dispersion interaction between neutral atoms (molecules). In particular, there are two interaction mechanisms:
\begin{enumerate}
  \item The fluctuation field of one cell polarizes the neighbour one. Therefore, the second cell gains a certain value of induced multipole moments. The latter interact with multipole moments of the first cell. The average value of such an interaction is an analogue of a dispersion interaction between neutral atoms (molecules):
      \begin{equation}\label{eq:multipole-ind_multipole_general}
      \Phi_1 = \langle W_{\text{mim}} \rangle,
      \end{equation}
      where index ``mim'' denotes the interaction between multipole and the induced multipole.
  \item  The interaction of fluctuation multipole moments of both cells. The average value of such an interaction is determined in a different way:
      \begin{equation}\label{eq:multipole-multipole_general}
      \Phi_2 = -\beta \langle W_{\text{mm}}^2 \rangle,
      \end{equation}
      where $\beta = 1/kT$.
\end{enumerate}

It is easy to ascertain that both components lead to the attraction between {``plasma atoms'' (i.e., electroneutral spherical cells)}. The dominating contribution in both cases is determined by ``dipole--dipole'' interactions, decreasing as $1/R^6$ ($R$ is an average distance between the centers of two grains). At the same time, since the grains have charges of the same sign, repulsion forces arise at distances {$R < 2 r_{\text D}$}, where Coulomb repulsion between dust grains is not screened. This allows us to qualitatively  model the interaction between two cells (i.e., grains) with the potential
\begin{displaymath}
U(R) = U_{\text r}(R) + U_{\text a}(R),
\end{displaymath}
where $U_{\text r}(R)$ and $U_{\text a}(R)$ are the repulsive and the attractive parts of $U(R)$, respectively. The repulsive part of $U(R)$ is modelled by the combination of the hard-core potential (the radius of hard-core coincides with the radius of a grain $r_{\text p}$) and the potential of Coulomb repulsion at distances {$2r_{\text p} < R < 2 r_{\text D}$}:
\begin{displaymath}
U_{\text r}(R) =
    \begin{cases}
    \infty, & R\leqslant 2r_{\text p}\,;\\
    \displaystyle \frac{(Ze)^2}{R}, & {2r_{\text p} < R < 2 r_{\text D}}\,.
    \end{cases}
\end{displaymath}
The attractive part $U_{\text a}(R)$ is conditioned by the forces of an electric multipole interaction:
\begin{displaymath}
U_{\text a}(R) = \Phi_1(R) + \Phi_2(R).
\end{displaymath}

The proposed interaction potential can lead to the formation of ordered structures for certain values of temperature and dust grain density. It should be noted that the random distribution of dust grains in the volume is not taken into account. Thus, the proposed approach could be applied to the regular homogeneous grain distribution (like in the case of dust crystals).

Let us determine the dominating contributions to $\Phi_1(R)$ and $\Phi_2(R)$ according to \eqref{eq:multipole-ind_multipole_general} and \eqref{eq:multipole-multipole_general}.

\subsection{``Dispersion'' intergrain interaction}
The average energy of electrostatic ``multipole--induced multipole'' interaction is determined according to \eqref{eq:multipole-ind_multipole_general} as follows:
\begin{equation}\label{eq:multipole-ind_multipole_general_2}
\Phi_1(R) = \langle W_{\text{did}} \rangle + \langle W_{\text{diq}} \rangle + \langle W_{\text{qiq}} \rangle + \ldots,
\end{equation}
where contributions $\langle W_{\text{did}} \rangle$, $\langle W_{\text{diq}} \rangle$ and $\langle W_{\text{qiq}} \rangle$ describe the averaged ``dipole--induced dipole'', ``dipole--induced quadrupole'' and ``quadrupole--induced quadrupole'' interactions, respectively.

The main contribution $\langle W_{\text{did}} \rangle$ can be estimated from the following simple considerations. It was noted above that the fluctuation field of one cell polarizes the other one. Therefore, the second cell acquires the induced dipole moment $\mathbf d^{(\text{ind})} = \alpha \mathbf d/R^3$, where $\alpha$ is the polarizability of a cell and $\mathbf d$ is the fluctuating dipole moment of the first cell. Thus, the energy of ``dipole--induced dipole'' interaction is $W_{\text{did}} = - \mathbf d \cdot \mathbf E^{(\text{ind})}$ (here $\mathbf E^{(\text{ind})} = \mathbf d^{(\text{ind})}/R^3$ is the induced field of the second cell). The careful analysis of this problem yields:
\begin{equation}\label{eq:dipole-ind_dipole_energy_general}
\Phi_1(R) = \langle W_{\text{did}} \rangle + \ldots \simeq - 4\alpha \frac{\langle \mathbf d^2 \rangle}{R^6}\,.
\end{equation}

\subsection{Electrostatic multipole intergrain interaction}
The average value of the direct electrostatic multipole interaction given by \eqref{eq:multipole-multipole_general} can be represented as follows:
\begin{equation}\label{eq:multipole-multipole_general_2}
\Phi_2(R) = -kT \left( \frac{A_6}{R^6} + \frac{A_8}{R^8} + \frac{A_{10}}{R^{10}} + \ldots \right).
\end{equation}
Here, coefficients $A_i$ ($i=6,\,8,\,10,\,\ldots$) are expressed through the average values of square multipole moments of a cell. To get the ``dipole--dipole'' contribution, we can follow the authors of work~\cite{kul_pre03} where the dipole fluid is considered. Thus,
\begin{equation}\label{eq:dipole-dipole_energy_general}
\Phi_2(R) \simeq - \frac23 \beta \frac{\langle \mathbf d^2 \rangle^2}{R^6}\,.
\end{equation}

The rest contributions to \eqref{eq:multipole-ind_multipole_general_2} and \eqref{eq:multipole-multipole_general_2} require a more detailed analysis and will be considered in a further paper.

\section{{Dipole} polarizability of a cell}\label{sec:4}

In the previous section, the model potential of intergrain interaction $U(R)$ has been introduced. This section is devoted to the {dipole} polarizability $\alpha$ of a cell. For this purpose, let us consider the reaction of electro-neutral cell to an external electric field $\mathbf E_0$. The latter polarizes a cell and it gains the polarization vector $\mathbf P = \alpha \mathbf E_0$. On the other hand, the polarization vector is a dipole moment of the unit cell volume:
\begin{equation}\label{eq:polarization_vector}
\mathbf P = \int \limits \mathbf r \cdot \rho(\mathbf r) \rd\mathbf r,
\end{equation}
where the volume-charge density $\rho(\mathbf r)$, as it was noted in section~\ref{sec:2}, is connected with the electrostatic potential $\phi(\mathbf r)$ by the Boltzmann distribution. In the presence of an external electric field (as opposed to the case considered above) the distributions of $\rho(\mathbf r)$ and $\phi(\mathbf r)$ lose their spherical symmetry: $\rho(\mathbf r)\Rightarrow\rho(r,\vartheta)$, $\phi(\mathbf r)\Rightarrow\phi(r,\vartheta)$.

The boundary conditions \eqref{eq:boundary_cond_1} for the Poisson equation have the following form:
\begin{equation}\label{eq:boundary_cond_2}
\left\{
\begin{aligned}
    &\phi(r_{\text c}, \vartheta) = - E_0 r_{\text c} \cos \vartheta,
\\
&    \left. \frac{\partial \phi(r,\vartheta)}{\partial r} \right|_{r = r_{\text p}} = - \frac{Z e}{r_{\text p}^2} + E_0 \cos \vartheta.
\end{aligned}
\right.
\end{equation}
Corresponding to \eqref{eq:boundary_cond_2}, the solution of the linearized Poisson equation in dimensionless variables \eqref{eq:dimless_varz} for the renormalized dimensionless potential $\psi(\tilde r,\vartheta) = 1 + e\phi(r,\vartheta)/kT$ has the form
\begin{equation}\label{eq:potential_psi}
\psi(\tilde r,\vartheta) = \psi_0(\tilde r) + \psi_{\text f}(\tilde r,\vartheta).
\end{equation}
Here, $\psi_0$ is the isotropic part of $\psi(\tilde r,\vartheta)$, determined by equation \eqref{eq:potential_psi_0}. The angular part $\psi_{\text f}$ (index ``f'' expresses the fact that this part of the potential is due to an external electric field) is proportional to $\cos\vartheta$ and is equal to
\begin{displaymath}
\begin{aligned}
\psi_{\text f}(\tilde r,\vartheta) = &-\tilde E_0 \cos \vartheta \left\{
    \frac{1}{\tilde r}
    \frac{\lambda \left( \varsigma\sinh\frac{\varsigma-\tilde r}{\lambda} - \lambda\cosh\frac{\varsigma-\tilde r}{\lambda} \right) + \varsigma^3 \left[ 2\lambda\sinh\frac{\tilde r - 1}{\lambda} + \left(2\lambda^2+1\right)\cosh\frac{\tilde r - 1}{\lambda} \right]}
         {\lambda\left( 2\varsigma-2\lambda^2-1 \right) \sinh\frac{\varsigma-1}{\lambda} + \left[ 2\lambda^2(\varsigma-1)+\varsigma \right]\cosh\frac{\varsigma-1}{\lambda}}
         \right.
    \nonumber\\
    &{}\left.+ \frac{1}{\tilde r^2}
    \frac{\lambda^2 \left( -\lambda\sinh\frac{\varsigma-\tilde r}{\lambda} + \varsigma\cosh\frac{\varsigma-\tilde r}{\lambda} \right) - \varsigma^3\lambda \left[ \left(2\lambda^2+1\right)\sinh\frac{\tilde r - 1}{\lambda} + 2\lambda\cosh\frac{\tilde r - 1}{\lambda} \right]}
         {\lambda\left( 2\varsigma-2\lambda^2-1 \right) \sinh\frac{\varsigma-1}{\lambda} + \left[ 2\lambda^2(\varsigma-1)+\varsigma \right]\cosh\frac{\varsigma-1}{\lambda}} \right\},
\label{eq:potential_psi_f}
\end{aligned}
\end{displaymath}
where $\tilde E_0 = eE_0r_{\text p}/kT$.

\begin{figure}[!h]
\centerline{
\includegraphics[width=0.65\textwidth]{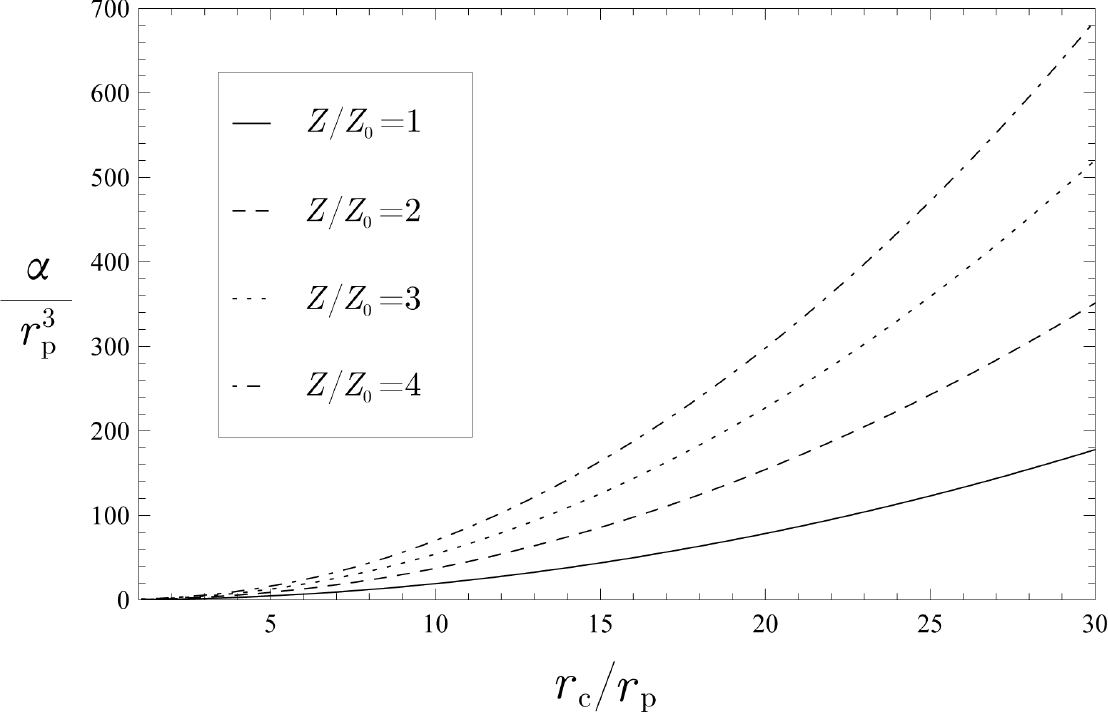}
}
\caption{Dimensionless polarizability $\alpha/r_{\text p}^3$ as the function of $r_{\text c}/r_{\text p}$.}
\label{fig:polarizability}
\end{figure}
The volume-charge density in \eqref{eq:polarization_vector} obviously connected with the potential $\psi(\tilde r, \vartheta)$ by the relation
\begin{displaymath}
\rho(r,\vartheta) = - \frac{kT}{4 \pi e r_{\text p}^2} \frac{1}{\lambda^2} \psi(\tilde r, \vartheta).
\end{displaymath}
Therefore, after integration and comparing the result with $P = \alpha E_0$, we obtain
\begin{eqnarray}
\alpha &\!\!=\!\!& \frac{r_{\text p}^3}{3\lambda} \left\{
    \frac{\varsigma^5 + \left\{ \varsigma^3 [ 2(\varsigma-3)\varsigma + 3 ] + 3\varsigma \right\}\lambda^2 + 3\left( 2\varsigma^3-1 \right)\lambda^4}
         {\lambda\left( 2\varsigma-2\lambda^2-1 \right) \sinh\frac{\varsigma-1}{\lambda} + \left[ 2(\varsigma-1)\lambda^2 + \varsigma \right] \cosh\frac{\varsigma-1}{\lambda}} \sinh\frac{\varsigma-1}{\lambda}\right.
    \nonumber\\
    &&{}\left.+\frac{(\varsigma-1) \left[ \varsigma^2(\varsigma-1)(2\varsigma+1) - \varsigma \right]\lambda - 3(2\varsigma^3-1)\lambda^2}
         {\lambda\left( 2\varsigma-2\lambda^2-1 \right) \sinh\frac{\varsigma-1}{\lambda} + \left[ 2(\varsigma-1)\lambda^2 + \varsigma \right] \cosh\frac{\varsigma-1}{\lambda}} \cosh\frac{\varsigma-1}{\lambda} \right\}.
\label{eq:polarizability}
\end{eqnarray}
The dependence of $\alpha/r_{\text p}^3$ on $r_{\text c}/r_{\text p}$ is shown in figure~\ref{fig:polarizability}. Note, that the value of Debye radius $\lambda$ for a corresponding $Z$ is determined from equation~\eqref{eq:neutral_eq_3}. Therefore, polarizability $\alpha$ also depends on the average dust grain charge $Z$. {Indeed, one can see that substitution of the potential \eqref{eq:potential_psi} into \eqref{eq:neutrality_condition} instead of \eqref{eq:potential_psi_0} also leads to \eqref{eq:neutral_eq_3}, so far as the integration over the angular part gives zero.}

\section{Fluctuation multipole moments of a cell}\label{sec:5}

In this section we introduce the method which allows one to construct the fluctuation electrostatic field inside and outside an electro-neutral cell. Moreover, a fluctuation dipole moment is obtained and the main contribution to the attractive part of inter-grain potential is presented in explicit form. {The fluctuation theory of electrostatic field for the case of ellipsoidal cells will be the subject of a separate~report.}

\subsection{Principles of construction of the fluctuation electrostatic field}

The electrostatic field deviates from its equilibrium value described in section~\ref{sec:2} due to fluctuations. Thus, let us consider two areas: inside (interior) and outside (exterior) a cell. We assume that the potentials of the fluctuation electrostatic field in the interior and exterior, $\phi'_{\text{in}}(\mathbf r)$ and $\phi'_{\text{ex}}(\mathbf r)$, respectively, satisfy the linear equations of the same type as the averaged potential $\phi_0(r)$:
\begin{equation}\label{eq:fluct_pois_in}
    \Delta \phi'_{\text{in}} (\mathbf r) - \frac{1}{r_{\text D}^2} \phi'_{\text{in}} (\mathbf r) = 0
\end{equation}
and
\begin{equation}\label{eq:fluct_laplace_ex}
    \Delta \phi'_{\text{ex}} (\mathbf r) = 0.
\end{equation}
It is supposed that the Laplace equation is proper in the exterior, as far as the whole dust grain charge is in a cell.

According to the form of equations \eqref{eq:fluct_pois_in} and \eqref{eq:fluct_laplace_ex}, we claim that potentials $\psi'_{\text{in}} \equiv e\phi'_{\text{in}}/kT$ and $\psi'_{\text{ex}} \equiv e\phi'_{\text{ex}}/kT$ have the following structure:
\begin{equation}\label{eq:fluct_psi_in}
\psi'_{\text{in}} (\tilde r, \vartheta, \varphi) = \sqrt{\frac{\pi\lambda}{2 \tilde r}} \sum\limits_{n=0}^{\infty} \sum\limits_{m=-n}^{n}
Y_{nm}(\vartheta, \varphi) \left[ C_{nm}^{(1)} I_{n + \frac12}(\tilde r / \lambda) + C_{nm}^{(2)} I_{-n - \frac12}(\tilde r / \lambda) \right]
\end{equation}
(here, $C_{nm}^{(1)}$ and $C_{nm}^{(2)}$ are the unknown coefficients, $I_{n + \frac12}(x)$ and $I_{-n - \frac12}(x)$ are the Modified Spherical Bessel functions of the first and the second kind, respectively) and
\begin{equation}\label{eq:fluct_psi_ex}
\psi'_{\text{ex}}(\tilde r, \vartheta, \varphi) = \sum\limits_{n=0}^{\infty} \sum\limits_{m=-n}^{n} \frac{\tilde D_{nm}}{\tilde r^{n+1}} Y_{nm} ( \vartheta, \varphi)
\end{equation}
(here, coefficients $\tilde D_{nm} = eD_{nm}/kTr_{\text p}^{n+1}$ are the required multipole moments of a cell). Here, we have used the dimensionless variables \eqref{eq:dimless_varz}.

The interconnection between the coefficients $C_{nm}^{(1)}$, $C_{nm}^{(2)}$ and $\tilde D_{nm}$ is determined from the conditions for continuity of potential and strength of the fluctuation electric field on the surface of a cell (see appendix~\ref{sec:appendix_a}):
\begin{equation}\label{eq:contin_cond}
\left\{
\begin{aligned}
    & \psi'_{\text{in}} (\varsigma,\vartheta,\varphi) = \psi'_{\text{ex}} (\varsigma,\vartheta,\varphi),
\\
  &  \left. \frac{\partial \psi'_{\text{in}} (\tilde r,\vartheta,\varphi)}{\partial \tilde r} \right|_{\tilde r = \varsigma} =
    \left. \frac{\partial \psi'_{\text{ex}} (\tilde r,\vartheta,\varphi)}{\partial \tilde r} \right|_{\tilde r = \varsigma}.
\end{aligned}
\right.
\end{equation}

According to the thermodynamic fluctuation theory \cite{LandLif5}, the average value of square multipole moments of a cell is
\begin{equation}\label{eq:average_mu_square}
\langle |D_{n0}|^2 \rangle \simeq \frac{kT}{\chi_n}\,,
\end{equation}
where $\chi_n$ are the coefficients in the expansion of fluctuation electrostatic field energy $W'_{\text{el}}$:
\begin{equation}\label{eq:fluct_energy_general}
W'_{\text{el}} = \sum\limits_{n=0}^{\infty} \chi_n |D_{n0}|^2.
\end{equation}
The explicit form of $W'_{\text{el}}$ is discussed in appendix~\ref{sec:appendix_b}. Thus, the comparison of equations \eqref{eq:fluct_energy_general} and \eqref{eq:fluct_energy_explicit} immediately gives us
\begin{equation}\label{eq:chi_n}
\chi_n = - \frac{1}{8\pi} \frac{1}{r_{\text p}^{2n+1}}
\left\{ \left[ L_i(n) a_i(1,n) \right] \cdot \left[ L_j(n) b_j(1,n) \right] +
\frac{1}{\lambda^2} \int\limits_{1}^{\varsigma} \left[ L_i(n) a_i(\tilde r,n) \right]^2 \tilde r^2 \rd \tilde r \right\},
\end{equation}
where the coefficients $a_i(1,n)$, $b_j(1,n)$, $L_i(n)$ and $L_j(n)$ ($i,j=1,2$) are determined by equations \eqref{eq:coeff_ab} and \eqref{eq:coeff_c}.

\subsection{Fluctuation dipole moment of a cell}

The dipole moment $\mathbf d$ of a cell required to determine the dominating contributions to the intergrain potential (see section~\ref{sec:3}) is obtained after setting in the expression for multipole moments $n=1$, i.e., $|\mathbf d| = D_{10}$. Therefore, the expressions \eqref{eq:coeff_ab} for coefficients $a_1(\tilde r, n)$, $a_2(\tilde r, n)$, $b_1(\tilde r, n)$ and $b_2(\tilde r, n)$ read as follows:
\begin{displaymath}
\begin{aligned}
a_1(\tilde r, 1) &= - \frac{\sinh\tilde r/\lambda}{(\tilde r/\lambda)^2} + \frac{\cosh\tilde r/\lambda}{\tilde r/\lambda}\,, \qquad
a_2(\tilde r, 1) = \frac{\sinh\tilde r/\lambda}{\tilde r/\lambda} - \frac{\cosh\tilde r/\lambda}{(\tilde r/\lambda)^2}\,,
\\
b_1(\tilde r, 1) &= \left( \frac{2\lambda^2}{\tilde r^3} + \frac{1}{\tilde r} \right) \sinh\tilde r/\lambda - \frac{2\lambda}{\tilde r^2} \cosh\tilde r/\lambda,
\\
b_2(\tilde r, 1) &= - \frac{2\lambda}{\tilde r^2} \sinh\tilde r/\lambda + \left( \frac{2\lambda^2}{\tilde r^3} + \frac{1}{\tilde r} \right) \cosh\tilde r/\lambda,
\end{aligned}
\end{displaymath}
\begin{figure}[!h]
\centerline{
\includegraphics[width=0.65\textwidth]{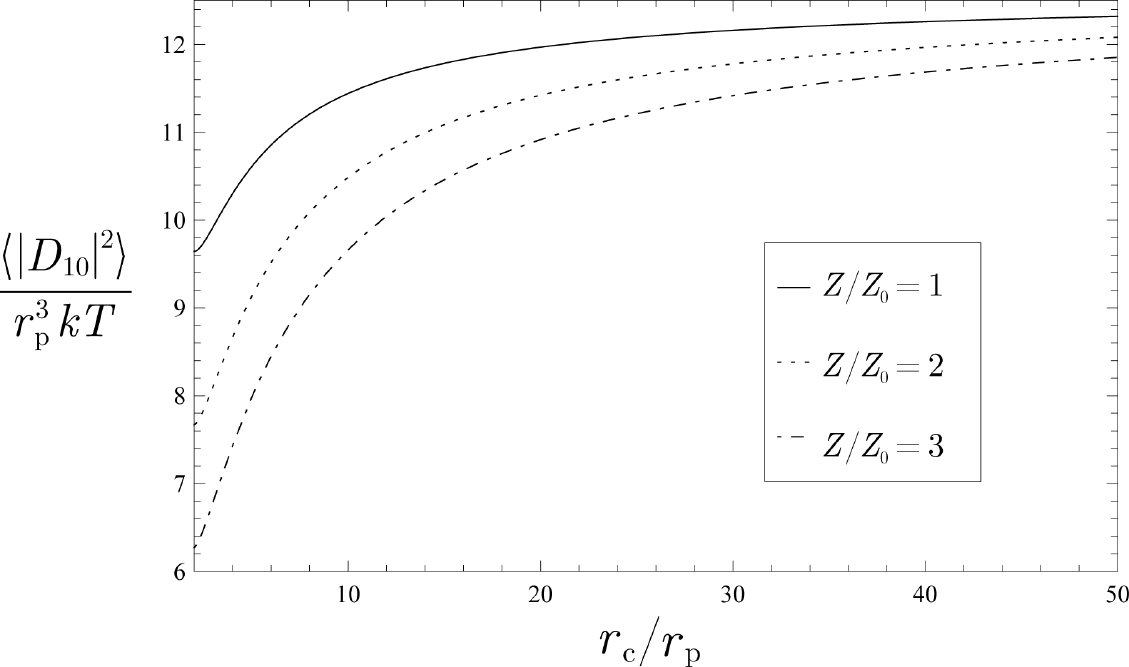}
}
\caption{The dependence of the fluctuation dipole momentum on the dimensionless radius of a cell $r_{\text c}/r_{\text p}$.}
\label{fig:dipole_moment}
\end{figure}
and for the mean-square value of a fluctuation dipole moment of a cell, after integration in \eqref{eq:chi_n} and after a simple calculation, we have
\begin{align}
    \langle |D_{10}|^2 \rangle = 16\pi \varsigma^2\lambda^2 \, r_{\text p}^3 kT\, \left[
    \frac12\lambda(8\lambda^2+1) \sinh 2\frac{\varsigma-1}{\lambda} + 2\lambda^2 (\lambda^2+1) \cosh 2\frac{\varsigma - 1}{\lambda} - 2\lambda^2(\lambda^2-1) - \varsigma+1
    \right]^{-1}.
    \label{eq:dipole_moment_square}
\end{align}
Let us consider the result obtained.

The mean-square value of a fluctuating dipole moment decreases when the dust grain charge increases (see figure~\ref{fig:dipole_moment}). Note, that this result is correct for  charges $Z/Z_0$ that are not very small. This limit is due to a general restriction on our approach: {$r_{\text D} < r_{\text c}$}. The case {$r_{\text D} > r_{\text c}$} corresponds to a rarefied gas, and thus the distribution functions for low-density plasma should be used.

\begin{figure}[h]
\begin{center}
{\resizebox{1.0\columnwidth}{!}{
\includegraphics[scale=1]{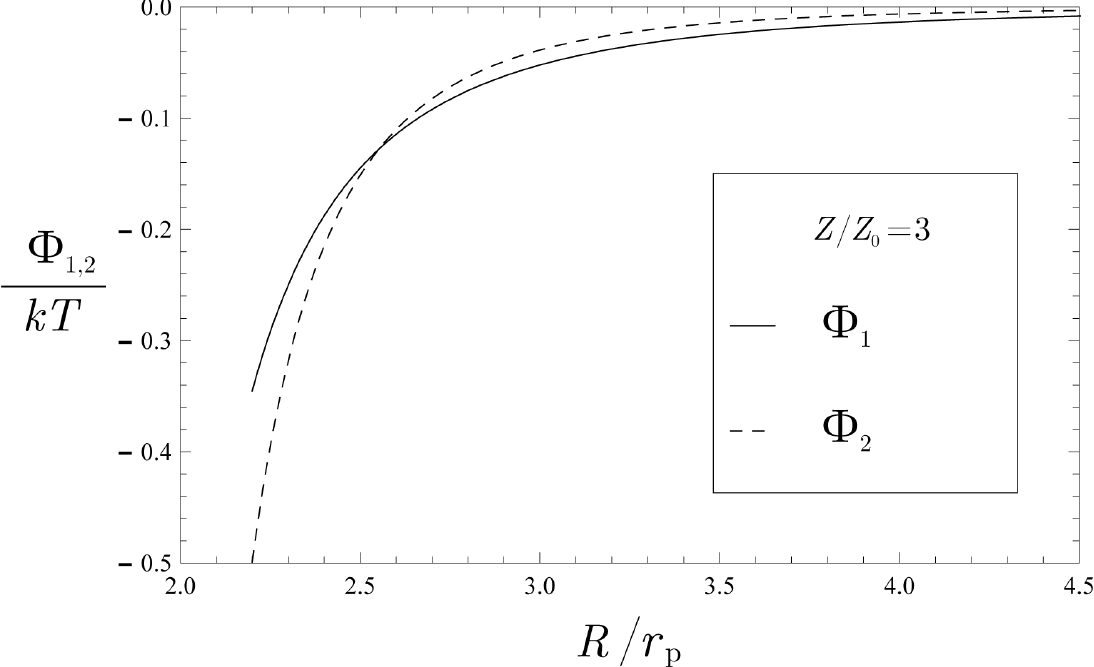}
\includegraphics[scale=1]{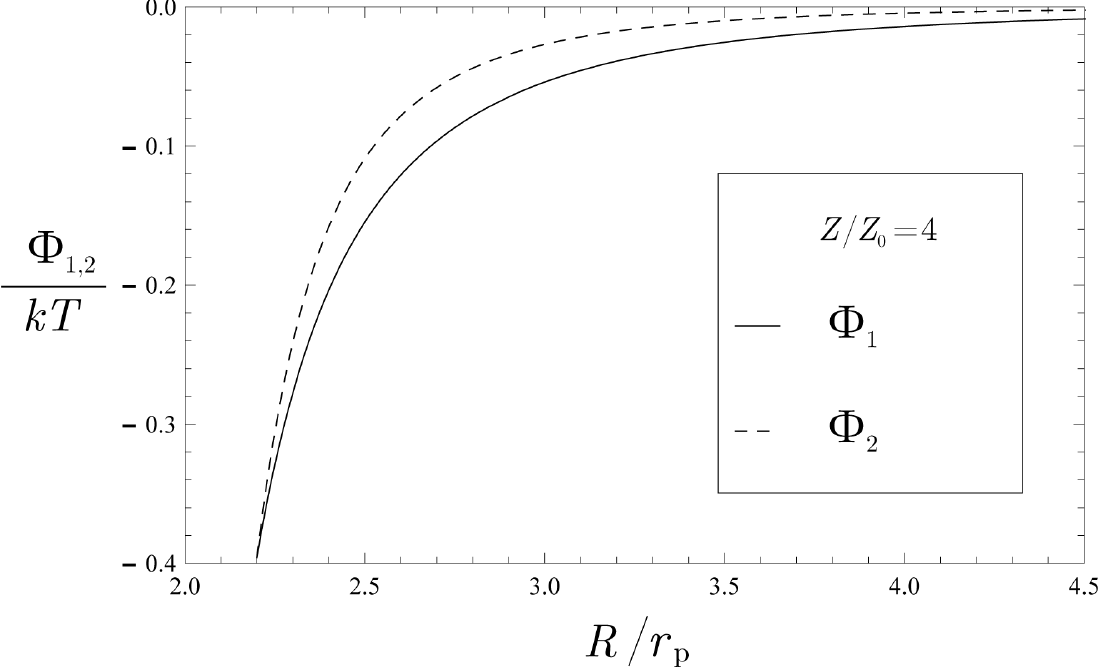}
}
}
    \end{center}
\caption{The dependencies of $\Phi_1/kT$ and $\Phi_2/kT$ on the average intergrain distance $R/r_{\text p}$ (in units of the radius of a grain). \emph{Left}: the case of $Z/Z_0=3$. \emph{Right}: the case of $Z/Z_0=4$.}
\label{fig:potentials}
\end{figure}
The dipole-dipole contributions to the attractive part of interaction potential [equations \eqref{eq:dipole-dipole_energy_general} and \eqref{eq:dipole-ind_dipole_energy_general}] are presented in figure~\ref{fig:potentials}. One can see that for small charges, the contribution $\Phi_2$ exceeds the contribution $\Phi_1$ at small intergrain distances. With an increase of the dust grain charge, the polarizability of a cell \eqref{eq:polarizability} increases (see figure~\ref{fig:polarizability}), and the contributions $\Phi_1$ and $\Phi_2$ become comparable. These contributions together, $\Phi_1 + \Phi_2$, lead to the formation of a potential well having the depth $\sim kT$.

\section{Conclusion}\label{sec:6}
The main results and conclusions of this paper are as follows.
\begin{itemize}
\item[(i)] The nature of long-range interaction between dust grains in complex plasma is discussed. It is supposed that plasma is a combination of electro-neutral cells of equal radius (mean-field approximation).
\item[(ii)] The main ideas of the fluctuation theory for electrostatic field in a cell model are formulated. The general expressions for fluctuation multipole moments are obtained. The average value of square fluctuation dipole moment is presented.
\item[(iii)] It is shown that the contributions of direct ``dipole--dipole'' interaction and ``dipole--induced dipole'' interaction form the potential well having the depth $\sim kT$. Multipole contributions of higher order lead to a further deepening of this well.
\end{itemize}

In molecular liquids, the attraction is determined by dispersion forces which decrease as $1/R^6$. The interaction between electro-neutral cells is similar to that in liquids. Thus, we expect that the grains in complex plasma form quasi-crystal structures like the molecules of argon in the transition to a solid state. The mechanism of ordering for solid argon is scrupulously considered in work~\cite{bondarev}. A detailed analysis of the proposed potential will be the subject of a further work.

{It is necessary to note that ``plasma atom'' is not fully similar to electroneutral molecules: it is more ``soft''. Thus, the use of a mean field approximation is incorrect for some problems. The characteristic example here is a problem of charge oscillations in dusty plasma. To determine the interaction potential between ``plasma atoms'', we consider charge fluctuations which  is also out of the mean field approximation. However, these fluctuations are small compared to the mean field profile of a charge as for the case of charge fluctuations generating disperse forces in molecular liquids.}

\section*{Acknowledgements}
The general statement of the problem and idea of its solution belong to Prof.~N.\,P.~Malomuzh. I thank him for that and for the help in preparing the manuscript. I am also grateful to S.\,A.~Lantratov for numerous discussions and to B.\,I.~Lev for the discussion of the obtained results during ``XII Workshop and Awards for young researchers in the field of statistical physics and condensed matter theory'' (Lviv, 2012). {Finally, I highly appreciate the anonymous Referees for stimulating remarks.}

\appendix
\renewcommand{\theequation}{\Alph{section}.\arabic{equation}}
\section{Fluctuation electrostatic field inside a cell}\label{sec:appendix_a}
By substituting the expressions \eqref{eq:fluct_psi_in} and \eqref{eq:fluct_psi_ex} in the continuity conditions \eqref{eq:contin_cond}, we get the following combined equations:
\begin{displaymath}
\left\{
\begin{aligned}
a_1(\varsigma, n) C_{nm}^{(1)} + a_2(\varsigma, n) C_{nm}^{(2)} &= \frac{\tilde D_{nm}}{\varsigma^{n+1}}\,,
\\
b_1(\varsigma, n) C_{nm}^{(1)} + b_2(\varsigma, n) C_{nm}^{(2)} &= - (n+1) \frac{\tilde D_{nm}}{\varsigma^{n+2}}\,,
\end{aligned}
\right.
\end{displaymath}
where coefficients $a_1(\tilde r, n)$, $a_2(\tilde r, n)$, $b_1(\tilde r, n)$ and $b_2(\tilde r, n)$ read
\begin{equation}\label{eq:coeff_ab}
\left\{
\begin{aligned}
a_1(\tilde r, n) &= \sqrt{\frac{\pi\lambda}{2\tilde r}} I_{n+\frac12}(\tilde r/\lambda), \qquad
a_2(\tilde r, n) = \sqrt{\frac{\pi\lambda}{2\tilde r}} I_{-n-\frac12}(\tilde r/\lambda),
\\
b_1(\tilde r, n) &= \sqrt{\frac{\pi\lambda}{2\tilde r}} \left[ \frac{1}{\lambda} I_{n+\frac32}(\tilde r/\lambda) + \frac{n}{\tilde r} I_{n+\frac12}(\tilde r/\lambda) \right],
\\
b_2(\tilde r, n) &= \sqrt{\frac{\pi\lambda}{2\tilde r}} \left[ \frac{1}{\lambda} I_{-n-\frac32}(\tilde r/\lambda) + \frac{n}{\tilde r} I_{-n-\frac12}(\tilde r/\lambda) \right].
\end{aligned}
\right.
\end{equation}
For coefficients $C_{nm}^{(1)}$ and $C_{nm}^{(2)}$, we obtain
\begin{equation}\label{eq:coeff_c}
\left\{
\begin{aligned}
C_{nm}^{(1)} &= L_1(n) \tilde D_{nm}, & L_1(n) &= \frac{1}{\varsigma^{n+2}} \cdot\frac{\varsigma b_2(\varsigma,n) + (n+1)a_2(\varsigma,n)}{a_1(\varsigma,n)b_2(\varsigma,n) - a_2(\varsigma,n)b_1(\varsigma,n)}\,,
\\
C_{nm}^{(2)} &= L_2(n) \tilde D_{nm}, & L_2(n) &= -\frac{1}{\varsigma^{n+2}} \cdot\frac{\varsigma b_1(\varsigma,n) + (n+1)a_1(\varsigma,n)}{a_1(\varsigma,n)b_2(\varsigma,n) - a_2(\varsigma,n)b_1(\varsigma,n)}\,.
\end{aligned}
\right.
\end{equation}

Thus, the fluctuation potential in the interior \eqref{eq:fluct_psi_in} can be represented as follows:
\begin{equation}\label{eq:fluct_psi_in_2}
\psi'_{\text{in}} (\tilde r, \vartheta, \varphi) = \sum\limits_{n=0}^{\infty} \sum\limits_{m=-n}^{n} Y_{nm}(\vartheta, \varphi) \sum\limits_{i=1}^2 L_i(n) a_{i}(\tilde r,n) \tilde D_{nm}\,.
\end{equation}

\section{Energy of fluctuation electrostatic field}\label{sec:appendix_b}
The energy of fluctuation electrostatic field is determined either by a field in the interior or by a field in the exterior:
\begin{displaymath}
W'_{\text{el}} = \frac{1}{8\pi} \int\limits_{V_{\text{in}}} \left[ \boldsymbol{\nabla} \phi'_{\text{in}}(\mathbf r) \right]^2 \mathrm d\mathbf r + \frac{1}{8\pi} \int\limits_{V_{\text{ex}}} \left[ \boldsymbol{\nabla} \phi'_{\text{ex}}(\mathbf r) \right]^2 \mathrm d\mathbf r,
\end{displaymath}
where $V_{\text{in}}$ and $V_{\text{ex}}$ are, respectively, the volume occupied by the electrons inside a cell and by the volume outside a cell.

Sequentially using  the transformation $(\boldsymbol{\nabla} \phi)^2 = \boldsymbol{\nabla} (\phi\boldsymbol{\nabla}\phi) - \phi\Delta\phi$, equation \eqref{eq:fluct_laplace_ex} and continuity conditions \eqref{eq:contin_cond}, we obtain
\begin{displaymath}
W'_{\text{el}} = - \frac{1}{8\pi} \left[ \oint\limits_{r=r_{\text p}} \phi'_{\text{in}} \frac{\partial \phi'_{\text{in}}}{\partial r} r^2 \rd\Omega +
\frac{1}{r_{\text D}^2} \int\limits_{r_{\text p}}^{r_{\text c}} (\phi'_{\text{in}})^2 r^2\rd r \right],
\end{displaymath}
or, in dimensionless variables \eqref{eq:dimless_varz} and using the explicit form \eqref{eq:fluct_psi_in_2},
\begin{equation}\label{eq:fluct_energy_explicit}
W'_{\text{el}} = - \frac{1}{8\pi} \sum\limits_{n=0}^{\infty}
\left\{ \left[ L_i(n) a_i(1,n) \right] \cdot \left[ L_j(n) b_j(1,n) \right] +
\frac{1}{\lambda^2} \int\limits_{1}^{\varsigma} \left[ L_i(n) a_i(\tilde r,n) \right]^2 \tilde r^2 \rd \tilde r \right\} \frac{|D_{n0}|^2}{r_{\text p}^{2n+1}}\,.
\end{equation}
Here, the summation over indexes $i,j=1,2$ occurs.

\newpage
\ukrainianpart

\title{Далекосяжна взаємодія між пилинками в плазмі}
\author{Д.Ю. Мішаглі\refaddr{label1}}
%
%
  \author{Д.Ю. Мішаглі}
  \addresses{\addr{ad1}Кафедра теоретичної фізики, фізичний факультет, Одеський національний університет імені І.І.~Мечникова, вул.~Дворянська~2, Одеса 65026, Україна
\addr{ad2} Інститут теоретичної фізики, Університет Лейпцига, 04109 Лейпциг, Німеччина}

\makeukrtitle

\begin{abstract}
\tolerance=3000
\looseness=-1Обговорено природу далекосяжної взаємодії між пилинками в плазмі. Ґрунтуючись на комірковій моделі запиленої плазми, запропановано потенціал взаємодії між пилинками. Притягальну частину потенціалу взаємодії представлено у вигляді мультипольної взаємодії між двома електронейтральними комірками. Це дозволило нам провести аналогію з молекулярними рідинами, де притягання між молекулами визначається дисперсійними силами. Також сформульовано основні ідеї теорії флуктуацій електростатичного поля в межах коміркової моделі та обчислено головний внесок в притягальну частину потенціалу взаємодії.
\keywords запилена плазма, пилові кристали, коміркова модель, потенціал взаємодії, флуктуації
\end{abstract}

\end{document}